## CERN Courier

### CERN COURIER

Oct 25, 2011
## Farewell to the Tevatron

**Installation of the world's first superconducting synchrotron began at Fermilab 30 years ago, but now the Tevatron has finally seen its last beam. Roger Dixon looks back at the intriguing story of this pioneering machine.**

**Résumé**

*Adieu au Tevatron*

*L'installation du premier synchrotron supraconducteur du monde a commencé à Fermilab il y a 30 ans. Aujourd'hui, le Tevatron a fait circuler son dernier faisceau. Au cours de son exploitation, il a fourni des faisceaux pour cible fixe, ainsi que des faisceaux en collision, ce qui a amené de nombreuses découvertes. L'idée d'un synchrotron supraconducteur a commencé avec Robert Wilson en 1967, avant la mise en place de ce qui est maintenant Fermilab. Les travaux menant au développement de la machine ont commencé sérieusement en 1973 et, dix ans plus tard, le « doubleur d'énergie », comme on l'appelait, accélérait son premier faisceau. Robert Dixon nous raconte l'histoire de l'accélérateur qui a été, pendant de nombreuses années, le champion du monde des hautes énergies.*

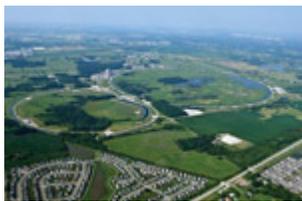

(http://images.iop.org/objects/ccr/cern/51/9/22/CCtev1_09_11.jpg)
Fermilab site (http://images.iop.org/objects/ccr/cern/51/9/22/CCtev1_09_11.jpg)

On 30 September 2011, Helen Edwards aborted the beam and dumped the ramp for the last time on what has for the past 28 years been one of the most productive physics machines in the world. The world's first superconducting particle accelerator represented a major advance in both technology and physics reach. The Tevatron's place in history is secure. During its life it provided fixed-target beams as well as colliding beams that resulted in numerous discoveries, including the first observations of the τ neutrino and top quark (**CERN Courier** October 2011 p20 (http://cerncourier.com/cws/article

/cern/47206)).

The concept of a superconducting accelerator predates the establishment of the National Accelerator Laboratory (NAL), later renamed Fermi National Accelerator Laboratory in 1974. In 1967 NAL's first director, Robert R Wilson, discussed the possibility of using superconducting technology soon after the new laboratory moved into temporary offices in Oakbrook, Illinois. He recognized that it was premature to begin developing the concept of a new machine before construction of the planned 500 GeV accelerator at NAL had even begun. Nevertheless, superconducting technology held the promise of higher energies and lower operating costs. Not only would a superconducting accelerator in the Main Ring tunnel double the energy of the fixed-target beams, it would also enable collisions between beams. The Intersecting Storage Rings at CERN had at that stage already proved the feasibility of colliding proton beams at 62 GeV in two conventional storage rings (**CERN Courier** January/February 2011 p27 (http://cerncourier.com/cws/article/cern/44855)). It would be a huge leap to go from conventional accelerator technology with one beam at NAL to a superconducting accelerator with colliding beams, but the thought was too tempting to dismiss completely.

### The superconducting challenge

The Main Ring was commissioned in 1972. It was completed under budget and on schedule even though many difficult problems were encountered – and then resolved – during construction. The laboratory's staff had demonstrated a desire to persevere and clearly had the talent to succeed in the face of tight budgets and enormous technical challenges. The Main Ring extended the energy reach by more than a factor of five over existing accelerators. The first 200 GeV beam to the fixed-target programme was a major accomplishment. Eventually, beams at 400 GeV with $3 \times 10^{13}$ protons per pulse were delivered and split between up to 15 experiments, resulting in many physics results, including the discovery of the Y in 1977.

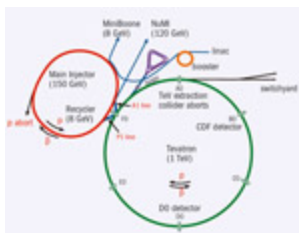
(http://images.iop.org/objects/ccr/cern/51/9/22/CCtev2_09_11.jpg)
Tevatron Run II (http://images.iop.org/objects/ccr/cern/51/9/22/CCtev2_09_11.jpg)

Once the Main Ring was commissioned the laboratory answered the call of the superconducting machine, initially known as the Energy Doubler/Saver because Wilson's vision was to reach an energy of 1000 GeV while also saving the cost of acceleration to lower energies. In 1973 work began in earnest to develop a superconducting accelerator magnet. Superconducting magnets had been built and used since the late 1940s and early 1950s – their primary use in particle physics being in bubble chambers. However, a new accelerator in the Main Ring tunnel would require approximately 1000 high-quality dipoles and quadrupoles: a reproducible magnet of accelerator quality would prove to be a major challenge.

Alvin Tollestrup played a key role in the effort to design such a magnet. After testing short magnets with monolithic superconductor, a design was chosen based on a warm-iron, collared coil of the niobium-titanium multifilament-strand cable developed at the Rutherford Laboratory in the UK. The first 20-ft (6.1-m) magnet was ready for tests in 1974 and by 1977 full-sized magnets were being produced and tested. However, many of these would be relegated to beam lines because further design improvements were still being implemented while magnet testing continued on test stands and in the beam lines.

An active quench-protection system had been developed and was exercised extensively during the early magnet testing phase – in which the people conducting the tests were ensconced behind the "dewar deflector". This experience led to a robust system that has worked well over the years.

**Towards construction**

In 1979, energy-deposition studies were carried out to measure the quench behaviour of two Energy Doubler dipoles in 350 GeV and 400 GeV beam extracted from the Main Ring. These measurements provided an early opportunity to use the MARS Monte Carlo shower simulation software that Nikolai Mokhov wrote at the Institute of High Energy Physics, Protvino, in 1974 and is now widely used for many accelerator and beam-related applications. Mokhov began visiting Fermilab with MARS in 1979. He helped to collect the data from the tests and used his software to determine that a superconducting collider should be feasible. A fixed-target machine was more uncertain; the extraction system would have to have better loss properties than the extraction system from the Main Ring. Helen Edwards and Mike Harrison came to the rescue with a modified design that moved the electrostatic extraction septa halfway round the ring from the extraction point, while Curtis Crawford developed a way to make the wire planes in the electrostatic septa straighter, so as to reduce losses.

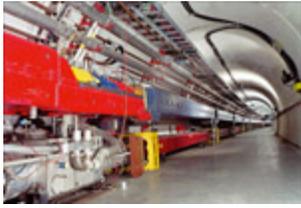
(http://images.iop.org/objects/ccr/cern/51/9/22/CCtev3_09_11.jpg)
Tevatron ring with red dipoles (http://images.iop.org/objects/ccr/cern/51/9/22/CCtev3_09_11.jpg)

Construction of the superconducting ring was authorized that same year and a final design for the magnets was in place by 1980. Because Wilson had anticipated building a second accelerator in the Main Ring tunnel, he left space underneath the Main Ring and designed its magnet stands to allow the magnets of a new machine to slip through them. The first step was to install magnets in one sector for a test in 1982. Concurrently, a large cryogenic refrigeration system was being built to provide the necessary cooling for the new accelerator. The cryogenic plant included 24 satellite refrigerators located in the service buildings that were spaced around the Main Ring tunnel. A large helium-liquefaction plant fed helium to the satellite refrigerators.

The completed accelerator was ready to be commissioned in 1983. It was a hectic and exciting time. Many challenges had been encountered and overcome, but many of those working on the project were still sceptical that it would succeed. Nevertheless, they made an incredible effort that brought the first superconducting synchrotron to life.

Beam was injected for the first time on 2 June 1983. It took less than a day to make the first turn all of the way round. On 3 July the Energy Doubler reached 512 GeV. Resonant extraction was established in August and the fixed-target programme at 400 GeV was underway in the autumn. By 1984 the energy had reached 800 GeV and the Energy Doubler was renamed the Tevatron. Five experiments took beam during the initial 400 GeV fixed-target run.

Construction of an antiproton source began in 1981, led by John Peoples. Antiprotons were stochastically cooled in the source using the technique that Simon van der Meer had first proposed at CERN (**CERN Courier** June 2011 p24 (http://cerncourier.com/cws/article/cern/46057)). Work also began to construct a collision hall in the BØ straight section that would accommodate the proposed CDF detector. The antiproton source was completed in 1985 and in October the first proton–antiproton collisions were observed in a partially complete CDF detector. The first collider-physics run began in 1987 using only the CDF detector. DØ came online in 1992 with a detector in the DØ straight section.

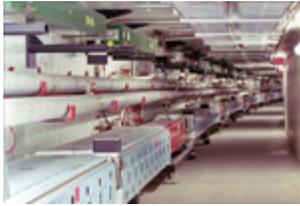

Main Injector ring (http://images.iop.org/objects/ccr/cern/51/9/22/CCtev4_09_11.jpg)

The Main Ring was still being used as an injector during the early collider runs, so it had to be accommodated in the collision halls. The CDF experiment had a Main Ring bypass that passed over the top of the detector, while the DØ collaboration had to learn to live with a Main Ring beam that went through the detector. In 1999 a new 150 GeV synchrotron, the Main Injector, was completed that replaced the Main Ring and provided more protons for both the collider and antiproton production. Built in a separate enclosure, it remedied the bypass problem. It would eventually enable simultaneous fixed-target and collider running, which had alternated until 1999.

In 1989 US President Bush awarded the National Medal of Technology to Helen Edwards, Rich Orr, Dick Lundy and Alvin Tollestrup for their work in building the Tevatron. Not only were they instrumental in solving the technical problems associated with building the forefront machine but they had also succeeded in maintaining an enthusiastic technical team in the face of problems that often seemed insurmountable.

## High luminosities

The design luminosity for the early running of the collider programme was $1 \times 10^{30}$ cm$^{-2}$s$^{-1}$ at 1800 GeV. During the first physics run in 1988 and 1989, $1.6 \times 10^{30}$ cm$^{-2}$s$^{-1}$ was achieved. By the end of Run I in 1996, initial luminosities were typically $1.6 \times 10^{31}$ cm$^{-2}$ s$^{-1}$ – a factor of 16 higher than the initial design luminosity. By this time a total integrated luminosity of 180 pb$^{-1}$ had been delivered to the two detectors – and the top quark had been discovered.

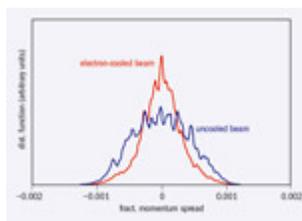

Energy spread (http://images.iop.org/objects/ccr/cern/51/9/22/CCtev5_09_11.jpg)

By 2001, when Run II began, many improvements to the accelerator complex had been made, including the addition of electrostatic separators to create helical orbits that prevented collisions at locations other than BØ and DØ, where the two detectors were situated. Antiproton cooling systems were also improved and the linac was upgraded from 200 MeV to 400 MeV to improve injection into the 8 GeV booster. Cold compressors were also added to the satellite refrigerators in 1993 to lower the operating temperature by 0.5 K, making it possible to raise the beam energy to 980 GeV. However, the new compressors were not used until the beginning of Run II in 2001.

The Main Injector had a larger aperture and could deliver more protons with higher efficiency. When Run II began, this enabled the delivery of more protons to the antiproton target and better transfer efficiencies for protons and antiprotons. There were also improvements to the Antiproton Source and the incorporation of a new permanent magnet ring, the Recycler, in the Main Injector tunnel. Initially meant to recycle antiprotons, it was never used for this purpose; instead it was used to stash and cool antiprotons delivered from the antiproton source.

Initial luminosities at the beginning of Run II were in the region of $2 \times 10^{31}$ cm$^{-2}$ s$^{-1}$. A luminosity improvement "campaign" was initiated and implemented concurrently with the physics programme. Improvements continued to be made over most of the Run II period. Significant improvements were made to the antiproton source resulting in an increase in the stacking rate from $7 \times 10^{10}$ to $26 \times 10^{10}$ antiprotons per hour. The Tevatron lattice was improved and magnets were reshimmed to correct problems with the "smart bolts". Slip stacking was developed in the Main Injector, which resulted in more protons on the antiproton target.

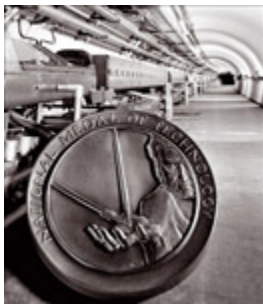
(http://images.iop.org/objects/ccr/cern/51/9/22/CCtev6_09_11.jpg)
National Medal of Technology (http://images.iop.org/objects/ccr/cern/51/9/22/CCtev6_09_11.jpg)

However, the largest single improvement made during Run II was the development and implementation of electron cooling in the Recycler. This effort, led by Sergei Nagaitsev, was commissioned in 2005 and resulted in smaller longitudinal emmitances. Using the Recycler to stash and cool also increased the stacking rate in the antiproton source because protons could be off-loaded to the recycler often, making the cooling more efficient. The net increase from electron cooling was more than a factor of two. Other improvements included a reduction of the $\beta^*$ in the two interaction regions and there was a vigorous programme to improve the reliability of the entire complex. Altogether the improvements resulted in initial luminosities a factor of 350 better than the original design.

During Run II, the Fermilab accelerator complex consisted of seven accelerators that together delivered beam for the collider programme, two neutrino beams and one test beam. It has performed magnificently over the years. All but the Tevatron will now continue operating to carry Fermilab into the future (**CERN Courier** September 2011 p54 (http://cerncourier.com/cws/article/cern/47216)). Nevertheless, the Tevatron defined the laboratory for 30 years. It has been an incredible experience for those of us fortunate enough to work on it.

### About the author

Roger L Dixon, Fermilab.